# Novel optically active lead-free relaxor ferroelectric $(Ba_{0.6}Bi_{0.2}Li_{0.2})TiO_3$

**Hitesh Borkar[1,2], Vaibhav Rao[1,3], Soma Dutta[4], Arun Barvat[1,2], Prabir Pal[1], M Tomar[5], Vinay Gupta[6], J. F. Scott[7], Ashok Kumar[1,2,]***

[1]CSIR-National Physical Laboratory, Dr. K. S. Krishnan Marg, New Delhi 110012, India
[2]Academy of Scientific and Innovative Research (AcSIR), CSIR-National Physical Laboratory (CSIR-NPL) Campus, Dr. K. S. Krishnan Road, New Delhi 110012, India
[3]Solar and Alternative Energy, Amity University, Jaipur, Rajasthan, 302006, India
[4]Materials Science Division, CSIR-National Aerospace Laboratories, Bangalore 560017, India.
[5]Department of Physics, Miranda House, University of Delhi, Delhi 110007, India
[6]Department of Physics and Astrophysics, University of Delhi, Delhi 110007, India
[7]Department of Chemistry and Department of Physics, University of St. Andrews, St. Andrews KY16 ST, United Kingdom

**Abstract:**

We discovered a near room temperature lead-free relaxor-ferroelectric $(Ba_{0.6}Bi_{0.2}Li_{0.2})TiO_3$ (BBLT) having A-site compositional disordered $ABO_3$ perovskite structure. Microstructure-property relations revealed that the chemical inhomogeneities and development of local polar nano regions (PNRs) are responsible for dielectric dispersion as a function of probe frequencies and temperatures. Rietveld analysis indicates mixed crystal structure with 80% tetragonal structure (space group *P4mm*) and 20% orthorhombic structure (space group *Amm2*) which is confirmed by the high resolution transmission electron diffraction pattern. Dielectric constant and tangent loss dispersion with and without illumination of light obey nonlinear Vogel-Future relation. It shows slim polarization-hysteresis (P-E) loops and excellent displacement coefficients ($d_{33}$ ~ 233 pm/V) near room temperature, which gradually diminish near the maximum dielectric dispersion temperature ($T_m$). The underlying physics for light-sensitive dielectric dispersion was probed by X-ray photon spectroscopy (XPS) which strongly suggests that mixed valence of bismuth ions, especially $Bi^{5+}$ ions, are responsible for most of the optically active centers. Ultraviolet photoemission measurements showed most of the Ti ions are in 4+ states and sit at the centers of the $TiO_6$ octahedra, which along with asymmetric hybridization between O 2*p* and Bi 6*s* orbitals appears to be the main driving force for net polarization. This BBLT material may open a new path for environmental friendly lead-free relaxor-ferroelectric research.

*Corresponding Author: Dr. Ashok Kumar (Email: ashok553@nplindia.org)*

## I. INTRODUCTION

Relaxor ferroelectrics (RFE) are one of the major subjects in solid state physics which deal with the physics of dynamic relaxation processes. During the past six decades scientists have been working to develop novel lead-free relaxor ferroelectrics with high electromechanical coefficients and high dielectric constants. Several physical models have been proposed, such as the random-field model, random-bond model, cationic order-disorder model, mean field model (spin glass and dipolar glass), polar nano-region model (PNRs), etc. [1-11] Relaxors are widely useful for technological applications such as transducers and actuators due to their exceptionally high dielectric constant, piezoelectric constant and electromechanical coefficients; however, optically active relaxor ferroelectrics may provide an extra degree of freedom for the development of power and energy efficient devices, such as modulators, tuning elements, optically active bistable switches, and memory elements. In the past several lead-based relaxors and lead-free relaxors have been invented; among them $PbMg_{1/3}Nb_{2/3}O_3$ (PMN) is a well known classical relaxor, and $PbSc_{1/2}Nb_{1/2}O_3$ (PSN) is a classical example of cation order-disorder at the B-site. [12-16] The $(1-x)(Na_{0.50}Bi_{0.50})TiO_3-xBaTiO_3$ (NBT-BT) compound near its morphotropic phase boundary (MPB) is one of the most highly investigated lead-free relaxor ferroelectrics with A-site cationic disorder.[17] $Ba(Zr,Ti)O_3$, $(Ba,Sr)TiO_3$ and $Ba(Sn,Ti)O_3$ are other popular lead-free materials with isovalent A/B-site cation substitution and have been explained on the basis of a soft pseudo-spin glass model. [18,19,20] The tetragonal tungsten bronze (TTBs) complex oxide structures include many relaxor ferroelectrics; however, the complex nature of their microstructure and microstructure-property relations restricts investigation on the basic mechanisms responsible for their relaxor behavior. [21] Li-doped $KTaO_3$ is also one of the most fascinating relaxors, showing dielectric dispersion due to dipolar interaction between randomly distributed off-center cations and a polarizable lattice. [22]

$LiNbO_3$ and $KNbO_3$ are well known electro-optic (EO) polar materials in which EO effects arise due to displacement of central cations with respect to oxygen anions in $NbO_6$ (B-site) octahedra and their interaction with incident light. In this situation an external light source falls on the crystal surface, which modifies the electronic band structure of the crystal and significantly changes the electrical response of the lattice. [23,24] $BaTiO_3$ has shown strong nonlinear electro-optic properties among the lead free perovskites. In the present investigation modification of $BaTiO_3$ by non-isovalent cations Li and Bi at the Ba-site leads to large changes in dielectric constant and dispersion under moderate laser light illumination. *Abel et al.* showed that $Si/SrTiO_3(4nm)/BaTiO_3$ (130 nm) heterostructures display at least five times larger Pockels coefficients of $r_{eff} = 148\,pm\,V^{-1}$, compared to standard material. [25] Actually two different unlikely mechanisms are involved in well known optically active high bandgap ferroelectric $BaTiO_3$ and low bandgap semiconducting ferroelectric SbSI ($T_c$ = 292 K, $E_g$ = 1.83 eV, and $P_s$ = 30 μC/cm$^2$), in the first system UV light changes $Ti^{+4}$ valence state whereas in SbSI illumination of light pumps electrons into the conduction band, respectively. The illumination of light also changes the phase transition temperature ($T_c$) of BaTiO$_3$ ($\Delta T_c$ = -2.6 K with 4700 Angstrom illumination) and affect the electronic structure of SbSI largely. [26,27,28]

The discovery of relaxor behavior in ABO$_3$ structures with B-site non-isovalent (PMN) and isovalent cations Ba(Zr,Ti)O$_3$, and Ba(Sn,Ti)O$_3$ led to the random field model and pseudo spin-glass model, respectively; however, it is still unclear as to what leads to relaxor behavior of non-isovalent substitution at A-sites of the ABO$_3$ structure. In this letter we have tried to understand these problems in addition of reporting the discovery of a novel optically active relaxor ferroelectric. We found that chemically unstable valence states of Bi ion may create an unstable chemical potential which is responsible for optically active centers, particularly the Bi$^{5+}$ ions. The symmetric lineshape of O1$s$ suggests high sample quality with stable ABO$_3$ structure. The photo-sensitivity of this compound is attributed to Bi valence fluctuations. This is a new and rather specific example of electronic ferroelectricity. [29,30]

## II.     SAMPLE PREPARATION AND EXPERIMENTAL METHODS

Polycrystalline (Ba$_{0.6}$Bi$_{0.2}$Li$_{0.2}$)TiO$_3$ (BBLT) ceramics were prepared by a conventional ceramic processing technique. Sigma-Aldrich high purity (~99.9%) precursor oxides and carbonates of Li$_2$CO$_3$, BaCO$_3$, Bi$_2$O$_3$, and TiO$_2$ were used to synthesize BBLT polycrystalline ceramics. Precursors for the desired stoichiometry were mechanically milled in wet conditions in IPA (isopropyl alcohol) for 2 h for homogeneous mixing. This mixed powder was calcined at 850 °C for 10 h. The granulated powder was shaped into circular discs of diameter 13 mm and thickness 1-2 mm under uniaxial pressure of 5–6 tons per square inch and later sintered in air at 1200 °C for 4 h to achieve nearly the maximum theoretical density. X-ray diffraction patterns were obtained using Bruker AXS D8 Advance X-ray diffractometer having the Cu-k$_\alpha$ (k$_\alpha$=1.5405A°) monochromatic radiation, in the 2θ range between 20° to 60°, and analyzed by Rietveld analysis. The temperature-dependent dielectric and polarization properties were measured by *Hioki LCR Meter model 3532-50* and *Radiant Precision Multiferroic Tester,* using a homemade probe station on as-grown samples. X-ray photoemission spectroscopy (XPS) measurements were performed by using an Omicron µ-metal ultra-high vacuum (UHV) system equipped with a monochromatic Al K$_\alpha$ X-ray source (hν=1486.7 eV) and a multi-channeltron hemispherical electron energy analyzer (EA 125). The ultraviolet photoemission spectroscopy (UPS) measurements were performed with a high-intensity vacuum ultraviolet source (HIS 13) at the He *I* (hν=21.2 eV) line, having a beam spot of 2.5 mm diameter. The total energy resolution was about 400 and 100 meV for monochromatic Al *K$\alpha$* and He *I* excitation, respectively. The voltage-displacement vector D(E) characteristic was obtained with an aixACCT Single Beam Laser Interferometer (aixSBLI). A scanning electron microscopy-energy dispersive X-ray (SEM-EDX) technique was used to ascertain the chemical compositions and grain distribution in the matrix using a Zeiss EVO MA-10. The determination of PNRs, crystal structures, and lattice planes was via a Technai G20, 300-kV, High Resolution Transmission Electron Microscope (HRTEM) on mechanically and ion-milled thin slabs.

## III. RESULTS AND DISCUSSION

### A. X-ray diffraction and Transmission Electron Microscopy

Fig1.(a) shows the Rietveld analysis of experimental data, which indicates the presence of both tetragonal phase (space group *P4mm*) and orthorhombic phase (space group *Amm2*) with volume ratio of 4:1 which is further confirmed by the HRTEM diffraction pattern (Fig. 1(b)). The fitting parameters *$\chi^2$ =1.72, Rp=7.26, Rw=9.88, and Rexp=7.53* for both systems suggest the accuracy of our assumptions for crystal structures those were well obtained within the fitting errors. SEM-EDX data revealed that the stoichiometry of the final composition matched with the initial assumption within +/- 10% error limit of the EDX. A SEM image in the inset of Fig.1 (a) indicates superior compactions of fine spherical grains with an average grain size less than a few microns. Chemically inhomogeneous dark patches with dimension less than 1 nm were seen throughout the matrix (marked in Fig1. (c)). Lattice planes observed by HRTEM matched with the data obtained from the (110) plane of XRD; however, it would be disingenuous to distinguish the lattice planes of both phases with low magnification HRTEM images, Fig1.(d). Interestingly, we were able to see the tiny PNRs with coexistence of both the phases; however, the tetragonal phase was the dominant phase in the matrix. It would also be difficult to distinguish the PNRs and chemically inhomogeneous regions. Chemical inhomogeneity, size of the grains, impurities, and crystal structures decide the sizes of PNRs and ferroelectric domains.

**B.     Dielectric Spectroscopy**

The dielectric constant and tangent loss of BBLT ceramics as a function of temperatures and frequencies under various conditions are shown in Fig. 2 (a-f). We have probed the dielectric response under three different warming conditions, depending on the experimental limitations: (i) cooling the system until 80 K and collecting the dielectric data on heating from 223 K to 673 K; (ii) dielectric data without cooling, i.e. from room temperature and above; and (iii) dielectric data from room temperature and above under illumination of weak 405-nm laser light with energy density 25 mW/cm$^2$. The important findings from the dielectric data are as follows: (i) significant relaxor shift in dielectric maximum temperature (from $T_m$ = 353 K at 1 kHz to 393 K at 1MHz); (ii) maximum tangent loss temperature shifts from 275 K at 1kHz to 302 K at 1 MHz and tangent loss values generally increases with increase in probe frequency -- a kind of finger print for relaxor ferroelectrics; (iii) the magnitude of $T_m$ values increases after cooling and illumination of external light and magnitudes of $T_m$ depend on probe frequencies; (iv) suppression of another phase transition near 523 K to 563 K, similar to the NBT-BT ferroelectric-to-paraelectric phase transition under illumination of light and finally, (v) significant increase in magnitude of dielectric constant ($\varepsilon$) at higher probe frequency under illumination of weak laser light (however the magnitude of tan $\delta$ remains nearly unchanged under illumination).

The increase in the dielectric constant and their dispersion can be explained on the basis of change in charge carrier concentrations (*n*). In general, for *n*-type ferroelectric semiconductor, electrons are responsible for change in dielectric constant behavior under illumination of light. This nonequilibrium charge carrier increases the concentration in the recharged levels (traps ~ N) and add an addition free energy function ($F_2$) to the total free energy of system. The net free energy of the crystal with an additional factor for electron subsystem can be written as;

$F = F_0+F_1+F_2$ ……………………………………………………………………………(1)

where $F_0(T)$ is free energy for paraelectric regions, $F_0(T)=F(P=0, \sigma_k=0, N_i=0)$,

$F_1$ is ferroelectric regions, $F_1 = F_0 + \frac{1}{2}\alpha P^2 + \frac{1}{4}\beta P^4$ ,………………………………………(2)

and electron subsystem (photo induced) $F_2 = \sum_i N_i E_i(T,P,\sigma_k) = N(E_g\text{-}u_1\text{-}u_2) = N\ddot{E}$…………..(3)

where $E_g$=bandgap, $u_1$ energy level of electron traps, $u_2$ energy level of hole traps, and a free-electron concentration *n* at energy level N, and $\ddot{E}$ corresponds to $E_g\text{-}u_1\text{-}u_2$. In present case, the contribution of recombination levels has been ignored assuming high resistance semiconductor; (for charge neutrality, p=n+N) where p and n represent free hole, and electron concentrations, respectively.

The energy term $\ddot{E}$ can be expanded in a series in P, T, and conductivity ($\sigma_k$) as follows:

$$\ddot{E}(P,T,\sigma_k) = \ddot{E}_0(T) + \frac{1}{2}\ddot{E}_P^{II}P^2 + \frac{1}{4}\ddot{E}_P^{IV}P4 + \sum_k \ddot{E}'_k \sigma_k + \frac{1}{2}\sum_k\sum_i \ddot{E}''_{ki}\sigma_k\sigma_i + P^2\sum_k \ddot{E}'''_k \sigma_k \quad\ldots\ldots(4)$$

where $\left(\ddot{E}_P^{II}\right)_0 = \left(\frac{\partial^2 \ddot{E}}{\partial P^2}\right) = a$ and $\left(\ddot{E}_P^{IV}\right)_0 = \left(\frac{\partial^4 \ddot{E}}{\partial P^4}\right) = b$, $\ddot{E}'_k = \left(\frac{\partial \ddot{E}}{\partial \sigma_k}\right)_0$, $\ddot{E}''_{ki} = \left(\frac{\partial^2 \ddot{E}}{\partial \sigma_k \partial \sigma_i}\right)_0$, and

$$\ddot{E}'''_{ki} = \left(\frac{\partial^3 \ddot{E}}{\partial P^2 \partial \sigma_k}\right)_0$$

Using the above mentioned equation 1 to 4, we obtain the free energy equation for the highly populated semiconductor ferroelectric under illumination of light:

$$F(T,P,N,\sigma_k) = F_{0N} + \frac{1}{2}\alpha_N P^2 + \frac{1}{4}\beta_N P^4 + N\sum_k \ddot{E}'_k \sigma_k - \frac{1}{2}\sum_i\sum_k s_{Nik}\sigma_i\sigma_k - P^2\sum_k v_{Nk}\sigma_k \quad\ldots\ldots(5)$$

where the coefficients are defined as $F_{0N} = F_0 + N\ddot{E}_0$, $\alpha_N = \alpha + aN$, $\beta_N = \beta + bN$, $v_{Nk} = v_k + \ddot{E}'''_k N$, and $s_{Nik} = s_{ik} + \ddot{E}''_{ik} N$

The change in free energy and wave function under illumination dictates the moderate change in dielectric constant, spontaneous polarization, piezoelectricity and shift in $T_c$ or $T_m$. All these parameters can be deduced from the free energy for ferroelectric semiconductor, equation 5.

After solving the free energy equation for second order ferroelectric phase transition, we found change in the Curie temperature under illumination of light is as follows:

$T_{0N}-T_0=\Delta T_N=-(C/2\pi)aN=-(\varepsilon_0 C)aN\ldots\ldots\ldots\ldots\ldots\ldots\ldots\ldots\ldots\ldots\ldots\ldots\ldots\ldots\ldots\ldots\ldots\ldots\ldots\ldots\ldots\ldots(6)$

C is Curie-Weiss constant, $T_{ON}$ and $T_0$ are Curie point with and without illumination of light. The condition for the free energy have minimum is $\left(\frac{\partial^2 F}{\partial P^2}\right)_0 = Na > 0$, it follows that a>0 and hence electrons lower the Curie point. N represents the electron concentration in traps near the conduction band. The relation between ε and C are as follows for ferroelectric regions (T<$T_0$): $\varepsilon = -C/[2(T-T_0)]$. Experimentally a small change in $\Delta T_N$ ~ 7 K at 10 kHz was observed which support the theoretical prediction of small change in $\Delta T_N$ under illumination in light.

The dielectric constant and temperature hysteresis ($\Delta T_h$) are related using the equation; $\varepsilon(T_c)\Delta T_h \cong C$, if we assume that electrons do not change the Curie-Weiss constant C, then they significantly reduce the temperature hysteresis $\Delta T_h \cong T_c - T_0 = \frac{1}{8\pi}C\frac{\beta^2}{\gamma}$ where $T_c$ phase transition temperature, $T_0$ is Curie-Weiss temperature, (α, β, γ) are constants, P is polarization and hence increase in dielectric constant $\varepsilon_N(T) > \varepsilon(T)$. [26-28] Experimentally a significant change in Δε ~ 30 at 10 kHz was observed which support the theoretical prediction of significant change in Δε under illumination in light.

## C. Vogel-Fulcher (VF) relation

The dielectric constant and loss maxima temperatures for various conditions were fitted with a nonlinear Vogel-Fulcher (VF) relation (Equation 1), as shown in Figure 3 [31,32,33].

$$f = f_0 \exp\left(-E_a / k_B(T_m - T_f)\right) \quad \text{...........(7)}$$

where $f$ is the experimental frequency; $f_0$, a pre-exponential factor corresponding to asymptotically high temperatures; $E_a$, the activation energy; $k_B$, Boltzmann's constant; and $T_f$, the static freezing temperature. The fitted parameters $E_a$ = 0.14 eV, 0.09 eV, 0.05 eV, and 0.07 eV (+/- 0.02 eV) and $f_0$ = 2.5x10$^{11}$ Hz, 3.5x10$^{12}$ Hz, 3.2x10$^9$ Hz, and 3.0x10$^{10}$ Hz and $T_f$ = 277 K, 242 K, 311 K, and 302 K (+/- 5 K) were obtained for the experimental data of $T_m$ with V-F fitting under condition of (i) measurement of dielectric constant and tangent loss data on cooled system, (ii) dielectric constant $T_m$ from 300 K (RT) without cooling, and (iii) dielectric constant $T_m$ from 300 K without cooling and under illumination of light, respectively. These values match with numerical values for other relaxor systems within the limit of their uncertainties [28]. For example, $f_o$ is typically of order 3-100 GHz.

### D. Polarization and Piezoelectric properties

Figure 4(a) & (b) show the polarization loops as a function of applied E-fields and temperatures without and with illumination light, respectively. In both the cases, a slim P-E hysteresis loop was observed near room temperature that drastically gets slimmer in the regions of dielectric dispersion window (RT to 450 K) and above which is nearly constant. Similar trends were observed for the saturation polarization ($P_s$) and coercive field ($E_c$) as shown in Fig. 4(c). This slim hysteresis near $T_m$ signifies classic relaxor systems. The monotonically large decrease in $P_s$ above $T_m$ suggests a possible strong candidate for electrocaloric effects.[34] The merger of $P_s$ near the $T_m$ supports the theoretical prediction of negligible effect of illumination of light on polarization magnitude near or above $T_m$. It can be clearly seen from figure 4 (a-c) that a significant change in magnitude of polarization was observed under light. The displacement-voltage (D-V) butterfly-like characteristic loops provide a low coercive field (~ 0.5 kV/cm) with saturation at 2 kV/cm. As shown in the Fig. 4(d), they prove that piezoelectric samples can be displaced mechanically with a coefficient of displacement ($d_{33}$) of about 233 pm/V. Note that our samples are thin ceramic slabs; in view of this, the $d_{33}$ values are quite comparable with lead- and lead-free ferroelectric thin films. [35] The low coercive field in displacement butterfly loops may be due to the experimental conditions where these experiments were performed: On a thin slab (~300 μm) at 10 Hz frequency compared to the 1.5 mm thick slab used for temperature dependent P-E measurements. The characteristic butterfly shape of the D-V plot due to a large bipolar signal is determined by the polarity reversal process in ferroelectrics when the coercive field strength is reached in the reversed field ($-E_c$). At the polarity reversal, the contraction of the sample reverses into expansion again, since the polarization and the field are again parallel. The slim feature of butterfly (D-V) and P-E hysteresis loops is a signature of classical relaxor ferroelectric material.

### E. X-ray photon spectroscopy

Room-temperature photoemission measurements were performed on *in-situ* heat treated samples at 623 K under a base vacuum of ~$5.0 \times 10^{-11}$ mbar. The heat treatment was performed *in-situ* for desorbing surface adsorbates. The binding energy of all the spectra was calibrated by measuring C 1*s* (285.0 eV) core levels which were monitoring throughout the experiment particularly before and after each core level spectra collected. The Bi 4*f* core-level XPS spectrum of BBLT is shown in Fig.5 (a). The spectra exhibit $4f_{7/2}$ and $4f_{5/2}$ spin-orbit doublet peaks located at ~157.8 (159.9) and ~163.2 (165.2) eV, respectively. The binding energy difference between the two pairs of spin-orbit peaks is about 5.4 (5.3) eV. A large separation of 2.1 eV between the two contributions of each spin-orbit photoemission signal was observed, suggesting that Bi is in a mixed valence state with $Bi^{3+}$ and $Bi^{5+}$ ion valency [36]. We performed double-peak fitting for 4$f_{7/2}$ and 4$f_{5/2}$ spectra, using a $\chi^2$ iterative program, and the concentration of $Bi^{3+}$ and $Bi^{5+}$ valence states is obtained about 55% and 45%, respectively. The Ba 3$d$ symmetric spectra are shown in Fig. 5 (b) with two pairs of spin–orbit split components at 778.4 (Ba 3$d_{5/2}$) and 793.6 (Ba 3$d_{3/2}$) eV, which indicates Ba ($Ba^{2+}$) present in +2 valence states. Fig. 5 (c) shows the highly symmetric (529.85 eV) O 1$s$ spectrum which is free from hydroxyl like species bound to different sites, normally observed at 532 eV.[37] It indicates a high-quality polycrystalline samples with minimal effect of grain boundaries in their functional properties. Figure 5(d) shows the Ti 2$p$ and Bi 4$d$ core-level spectrum. The Ti 2$p$ spectrum shows 2$p_{3/2}$ and 2$p_{1/2}$ spin-orbit doublet peaks located at ~ 458.6 and 464.0 eV, respectively. We have fitted the 2$p_{3/2}$ spectrum by two distinct features: a small signal at 456.5 eV attributed to the $Ti^{3+}$ (15%) component, and the $Ti^{4+}$ main signal (85 %) appearing at higher binding energy at 458.5 eV. The small $Ti^{3+}$ signal lies nearly withn the experimental fitting errors (+/- 5%) [38]. The Bi 4$d$ spectrum shows 4$d_{5/2}$ and 4$d_{3/2}$ spin-orbit doublet peaks located at ~ 440.0 and 463.5 eV, respectively. Figure 5(e) shows the core level spectra of Ti 3$s$ and Li 1$s$. The sharp peak of Ti 3$s$ shadowed the lattice site of Li cations generally mentioned at 55.6 eV; however, strong evidence of interstitial Li cations at 51.8 eV obtained, compared to the lattice occupancy of Li ions, from 53 to 55 eV, depending on the chemical potential of substitution site. [39,40] It was more difficult to attribute unambiguously the peak at 52 eV as the substitution site of Li cations.

The angle-integrated valence band photoemission spectrum of the sample is shown in Fig. 5(f). A major sharp feature at ~2.4 eV of $Ti^{4+}$ states of the $TiO_6$ octahedra was obtained; however, a very small feature at 1.5 eV was also seen due to the Ti 3$d^1$ states from $Ti^{3+}$ valence state (~ 5%). The features in the binding energy region of 2-3 eV are mainly from the hybridized Ti 3d ($Ti^{4+}$)-O 2p states, while the features between 3-7 eV are strongly mixed character of O 2$p$ and Bi 6*s* states with a small Ti 3*d* contribution. All together, XPS and UPS results confirm the presence of mixed valence states of Bi with polarization due to displacement of $Ti^{4+}$ cations from the center of octahedra and hybridization of Bi cations with neighboring oxygen 2*p* orbitals.

## IV. Conclusions

We have successfully developed a stable novel lead-free relaxor ferroelectric after partial (40%) substitution of Ba-site of $BaTiO_3$ with optically active Bi and Li cations. Bismuth exists in mixed valence states along with freely available Li cations at the interstitial sites, which make the BBLT more optically active. This $Li^{1+}$ and $Bi^{5+}$ cations act as an optically active center, which significantly enhances the dielectric constant under moderate illumination. The mixed crystal phases, chemically inhomogeneous nanoscale regions, PNRs, and mixed valence states provide a favorable condition for the dielectric dispersion, high piezoelectric coefficients, and development of optically active nano-domains. A thermally activated phase transition (onset of a paraelectric phase) near 523 K-563 K is significantly suppressed under illumination of weak light which might be a topic of further research. The frequency dependent phase transition temperatures follow a nonlinear dependence of probe frequencies and a Vogel-Fulcher relation. The presence of external light source drastically changes the dielectric dispersion. One can easily see that light does not allow the merger the dielectric spectra with the underlying conduction background which generally occurs after $T_m$ (under normal dark conditions). The dielectric dispersion persists well above the Burns temperature ($T_B$), and it suggests the presence of PNRs -- chemically and optically active regions for $T \gg T_m$ and perhaps also $T > T_B$; however, at elevated temperatures thermally activated charge carriers may also be responsible for large dielectric dispersion. As expected from relaxor ferroelectrics, we have obtained a high value of displacement coefficient of 233 pm/V, which makes BBLT more attractive for optically active actuators, sensors and bistable switches.

**Acknowledgement:**

AK acknowledges the CSIR-MIST (PSC-0111) project for their financial assistance. P. P. thanks CSIR network projects NWP55, PSC0109 and OLP120132 for financial support. Hitesh Borkar would like to acknowledge the CSIR (JRF) to provide fellowship to carry out Ph. D program. A.B. thanks UGC (JRF), India for fellowship towards his Ph. D program. Authors would like to thank Dr. V N Ojha (Head ALSIM) and Dr. Sanjay Yadav, for their constant encouragement.

**Figure captions**

Fig. 1 (a) shows XRD patterns and its Rietveld analysis fitting using mixed phase process i.e. tetragonal and orthorhombic crystal structures, SEM image for surface morphology is shown in inset. Fig 1 (b, c, & d) illustrate mixed phase diffraction patterns, existence of nanoscale chemical inhomogeneous regions, and lattice resoled (110) crystal planes, respectively. Nanoscale dark patches are marked in the Fig1.(c).

Fig. 2 (a-f). Dielectric constant and tangent loss spectra as functions of temperature and frequencies, (a-b) after cooling the system till 80 K and measurement from 223 K to 673 K, (c-d) in dark enclosure and above room temperature, and (e-f) under constant illumination of weak laser light with energy density 25 mW/cm$^2$. Arrow lines indicate the direction of increasing frequencies. In all three conditions, the magnitude of dielectric constant at $T_m$ decreases with increasing frequencies (from 1 kHz to 1MHz), however order was just opposite for tangent loss.

Fig. 3 Nonlinear VF relation was valid for all three conditions for both dielectric constant and tangent loss $T_m$ behavior.

Fig. 4 The temperature dependent slim P-E loops as function of temperatures and E-fields are presented, (a) in dark enclosure, (b) under illumination of light, (c) the variation of $2P_s$, and $2E_C$ with and without light as a function of temperatures to check the PNRs behavior near $T_m$, and (d) displacement-voltage (D-V) butterfly characteristic loops.

Fig. 5 Shows the XPS and UPS spectra (a) Bi 4$f$ core-level, (b) Bi 3$d$ spin –orbit split components, (c) O 1$s$ at 529.85 eV, (d) Ti 2$p$ spin-orbit doublet peaks, (e) Ti 3$s$ and Li 1$s$ core level spectra, and (e) Valence band photoemission spectrum taken using photons of energy of 21.2 eV.

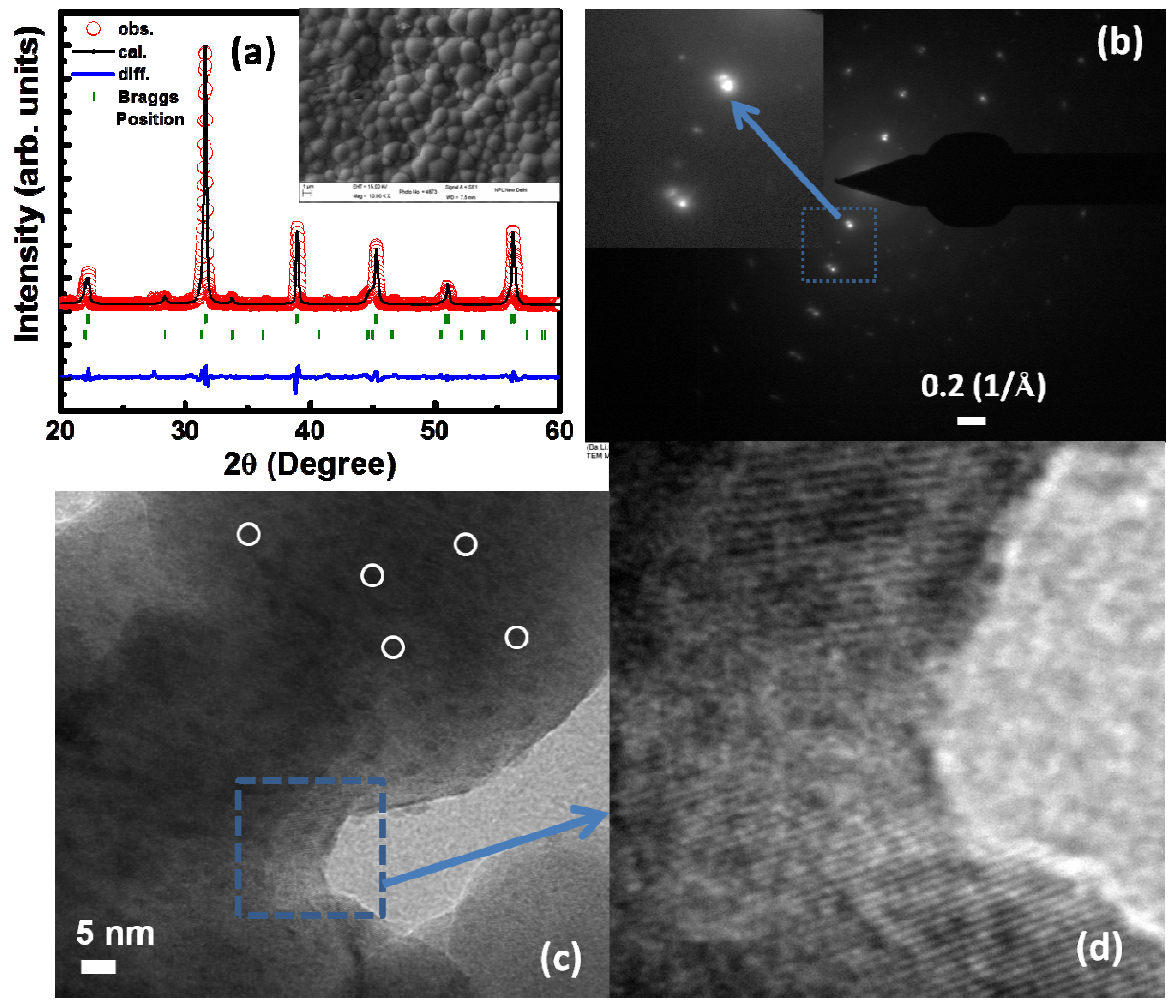

**Fig.1**

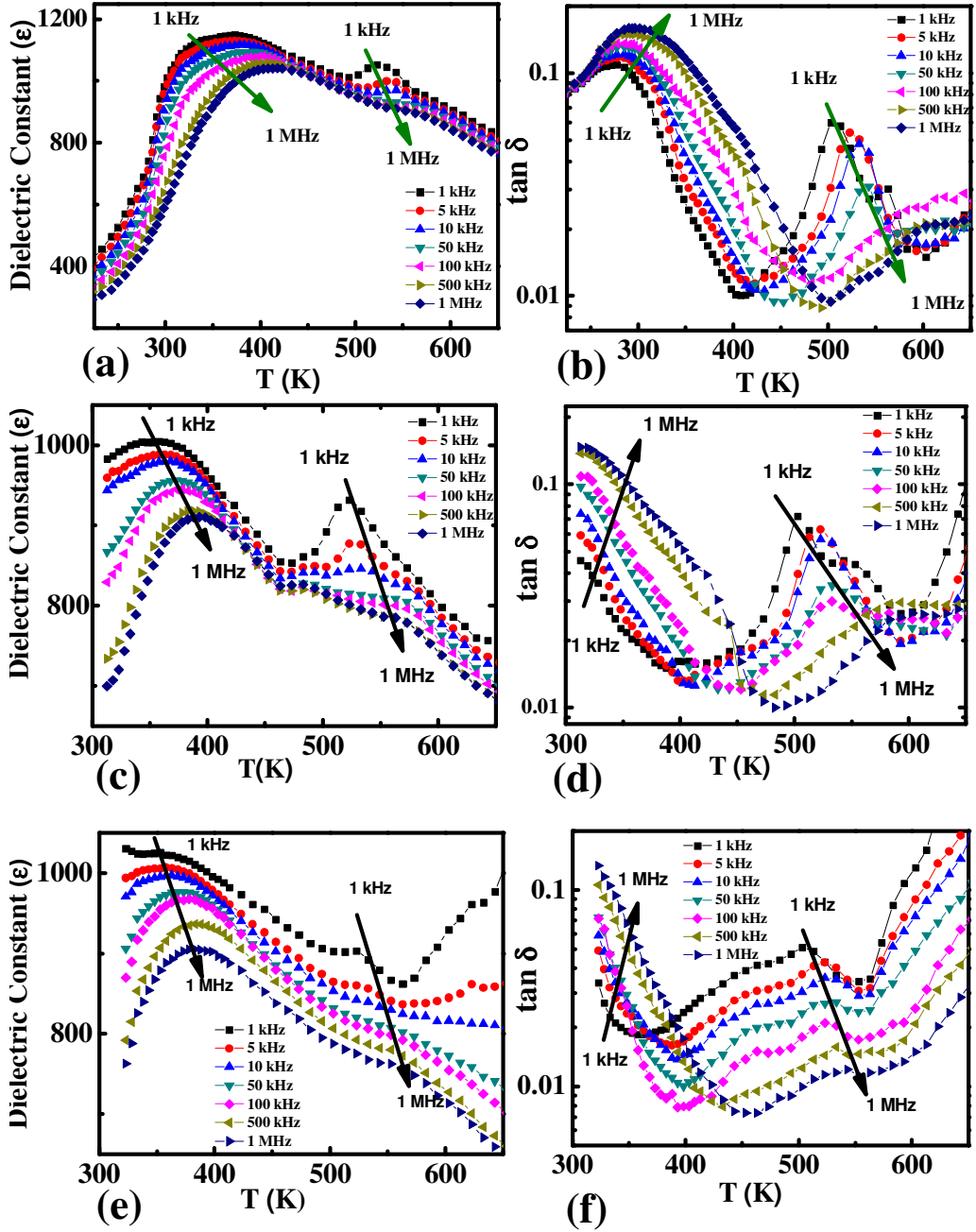

**Fig.2**

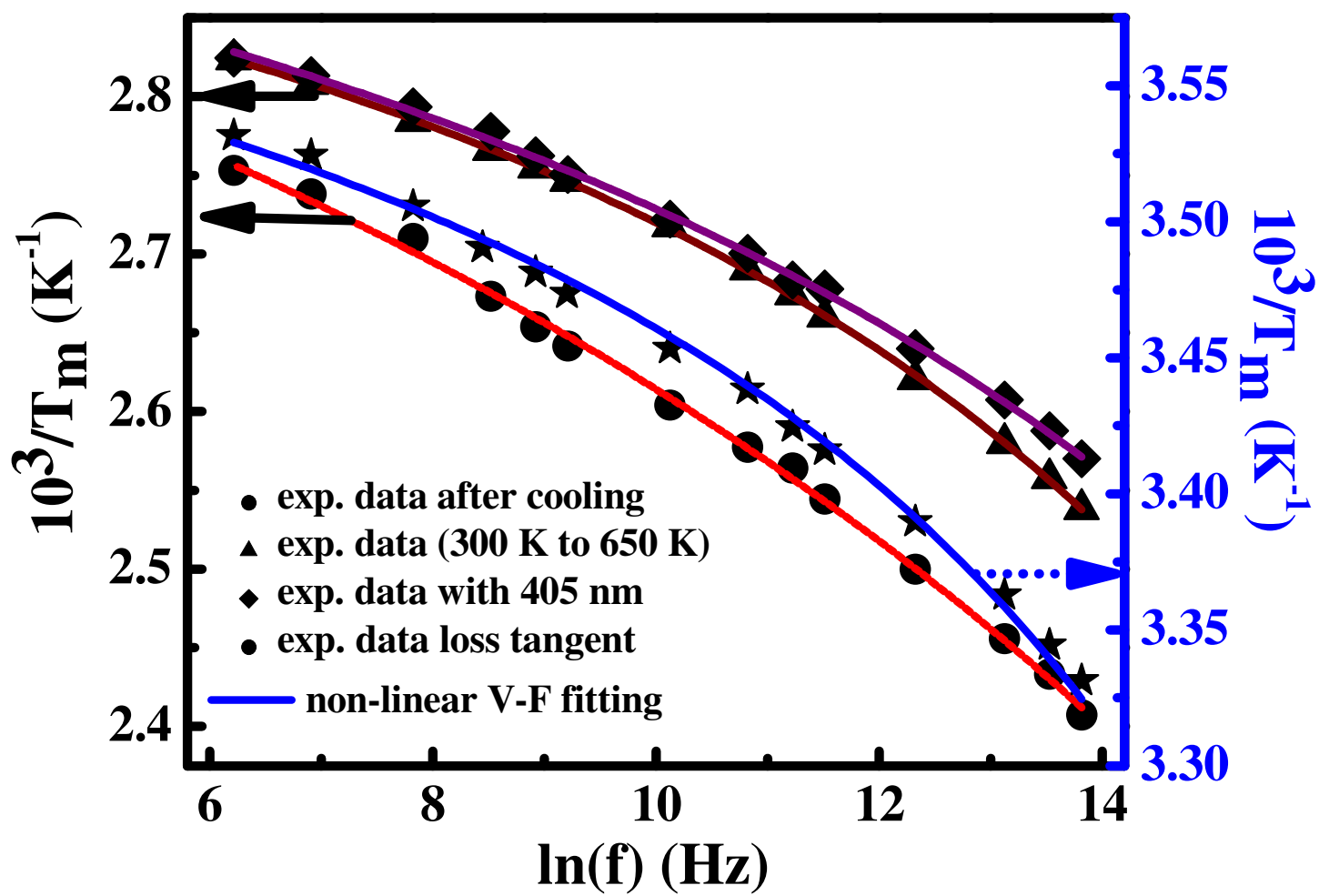

**Fig.3**

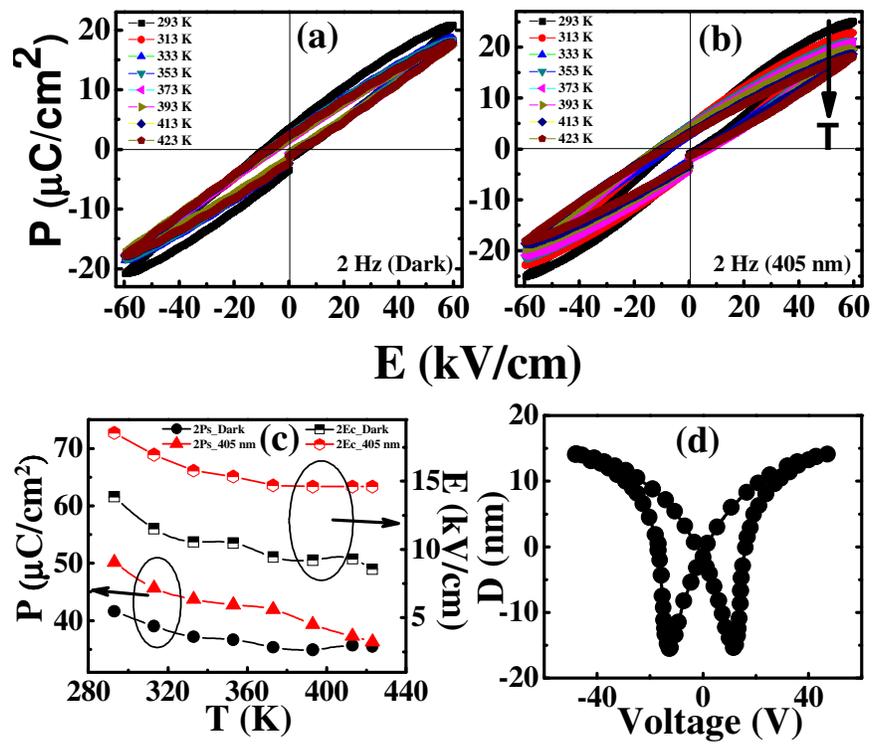

**Fig.4**

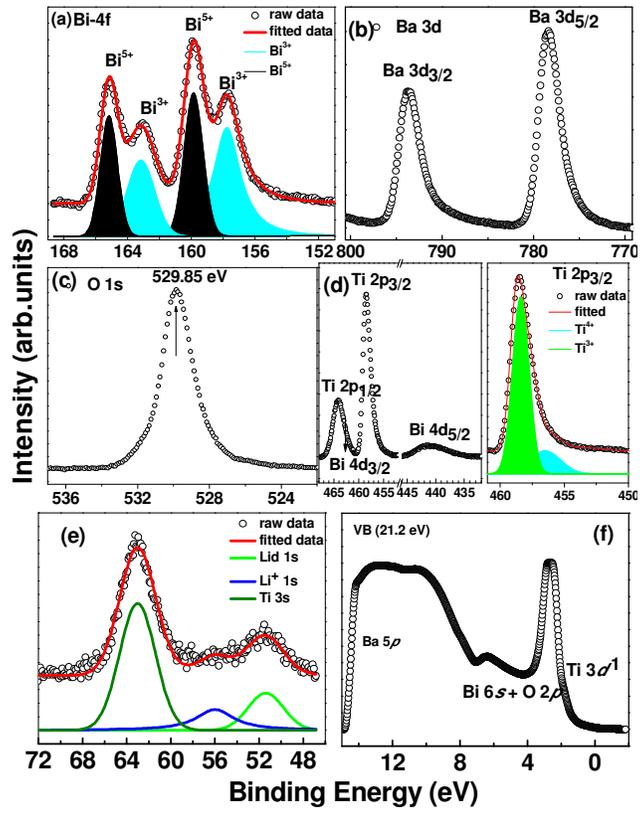

**Fig.5**

**References:**

[1] Blinc R., Laguta V. V., Zalar B., *Phys. Rev. Lett.* **2003,** 91, 247601-247604.

[2] a) Jeong I. K. et al., *Phys. Rev. Lett.* **2005,** 94, 147602-147604; b) Dkhil B. et al., *Phys. Rev. B,* **2001,** 65, 024104-024108; c) Gehring P. M., Wakimoto S., Ye Z. G., and Shirane G., *Phys. Rev. Lett.* **2001,** 87, 277601-277606.

[3] a) Vugmeister B. E., Rabitz H., *Phys. Rev. B* **1998,** 57, 7581-7585; b) Vugmeister B. E., *Phys. Rev. B* **2006,** 73, 174117-174127.

[4] S. Tinte, B. P. Burton, E. Cockayne, U.V. Waghmare, *Phys. Rev. Lett.* **2006,** 97, 137601-137604.

[5] Randall C. A., Bhalla A. S., *Jpn. J. Appl. Phys.*, **1990,** 29, 327-330.

[6] Bell A. J., *J. Phys.: Condens. Matter*, **1993,** 5, 8773-8792.

[7] Glazounov A. E., Bell A. J., Tagantsev A. K., *J. Phys.: Condens. Matter*, **1995,** 7, 4145-4168.

[8] Viehland D., Jang S. J., Cross L. E., Wuttig M., *J. Appl. Phys.*, **1990,** 68, 2916-2921.

[9] Glazounov A. E., Tagantsev A. K., *Phys. Rev. Lett.*, **2000,** 85, 2192-2195.

[10] Pirc R., Blinc R., *Phys. Rev. B*, 1999, **60**, 13470-13748.

[11] Westphal V., Kleemann W., Glinchuk M. D., *Phys. Rev. Lett.*, **1992,** 68, 847-850.

[12] Smolenskii G. A., Isupov V. A., Agranovskaya A. I., Popov S. N., *Sov. Phys. Solid State* **1961,** 2, 2584-2594.

[13] Burns G., Dacol F. H., *Solid State Commun.*, **1983,** 48, 853-856.

[14] Xu G., Shirane G., Copley J. R. D., Gehring P. M., *Phys. Rev. B*, **2004,** 69, 064112-064116.

[15] Fu D., Taniguchi H., Itoh M., Koshihara S., Yamamoto N., Mori S., *Phys. Rev. Lett.*, **2009,** 103, 207601-207604.

[16] Cowley R. A., Gvasaliya S. N., Lushnikov S. G., Roessli B., Rotaru G. M., *Adv. Phys.* **2011** 60, 229-327.

[17] Garg R., Rao B. N., Senyshyn A., Krishna P. S. R., Ranjan R., *Phys. Rev. B*, **2013,** 88, 014103-014115.

[18] Sherrington D., *Phys. Rev. Lett.*, **2013,** 111, 227601-227605.

[19] Rivera I., Kumar A., Ortega N., Katiyar R. S., Lushnikov S., *Solid State Communications*, **2009, 149 (3)**, 172-176.

[20] Maiti T., Guo R., Bhalla A. S., *J. Am. Ceram. Soc.*, **2008,** 91, 1769-1780.

[21] Zhu X., Fu M., Stennett M. C., Vilarinho P. M., Levin I., Randall C. A., Gardner J., Morrison F. D., Reaney I. M., *Chem. Mater.*, **2015,** 27(9), 3250-3261.

[22] Toulouse J., Vugmeister B. E., Pattnaik R., *Phys. Rev. Lett.*, **1994** 73, 3467-3470.

[23] Meng X. Y., Wang Z. Z., Zhu Y., Chen C. T., *J. Appl. Phys.*, **2007,** 101, 103506-103506.

[24] Chen C. T., Yang H. T., Wang Z. Z., Lin Z. S., *Chem. Phys. Lett.*, **2004,** 222, 397-400.

[25] Abel S., Stöferle T., Marchiori C., Rossel C., Rossell M. D., Erni R., Caimi D., Sousa M., Chelnokov A., Offrein B. J., Fompeyrine J., *Nature Communications*, **2013**, **4**,1671-1674.

[26] Fridkin V. M., Photoferroelectrics, *Springer-Verlag, Berlin*, **1979**.

[27] Fridkin V. M. et al. (photoresponse in ferroelectric SbSI): *Appl. Phys. Lett.*, **1967,** 10, 354-356.

[28] Belyaev M., Fridkin V. M., Grekov A. A., Kosonogov N. A., Rodin A. I., *J. de Physique Colloq.*, **1972,** 33, *C*2123-125.

[29] Yamauchi K. and Barone P., *J. Phys Cond Mat.*, **2014**, 26, 103201-103218.

[30] Jeng H. T., Ren C.Y, and Hsue C.S., *Phys. Rev. B*, **2011,** 84, 024421-024426.

[31] Vogel H., *Z. Phys*, **1921**, 22, 645-646.

[32] Tagantsev A. K., *Phys. Rev. Lett.*, **1994,** 72**,** 1100-1103.

[33] Correa M., Kumar A., Priya S., Katiyar R. S., Scott J. F., *Phys. Rev. B*, **2011,** 83, 014302-10.

[34] Mischenko A. S., Zhang Q., Scott J. F., Whatmore R. W., Mathur N. D., *Science*, **2006,** 311, 1270-1271.

[35] K. Rabe, Ch H. Ahn. J. M. Triscone (Eds.), *Physics of Ferroelectrics– A Modern Perspective, Springer,* (2007).

[36] Zalecki R., Woch W. M., Kowalik M., Kołodziejczyk A., Gritzner G., *Acta Phys. Pol. A*, **2010,** 118, 393-395.

[37] Kumar P., Pal P., Shukla A. K., Pulikkotil J. J., and Dogra A., *Phys. Rev. B*, **2015,** 91, 115127-115138.

[38] Pal P., Kumar P., Aswin V., Dogra A., Joshi A. G., *J. Appl. Phys.*, **2014,** 116, 053704-053706.

[39] Schroder K. W., Celio H., Webb L. J., Stevenson K. J., *J. Phys. Chem. C*, **2012,** 116, 19737-19747.

[40] Awan S. U., Hasanain S. K., Bertino M. F., Jaffari G. H., *J. Phys.: Condens. Matter*, **2013,** 25, 156005-156015.